\documentclass[twocolumn,tightenlines,showpacs,aps]{revtex4}
\usepackage{graphicx}

\begin{document}

\title{Plasmon-mediated Polarization Anomalies in Surface Enhanced Raman Scattering (SERS)}

\author{E. C. Le Ru} \email{Eric.LeRu@vuw.ac.nz}

\author{P. G. Etchegoin} \email{Pablo.Etchegoin@vuw.ac.nz}

\affiliation{The McDiarmid Institute for Advanced Materials and Nanotechnology\\
School of Chemical and Physical Sciences\\ Victoria University of Wellington\\
PO Box 600, Wellington, New Zealand}

\date{\today}

\begin{abstract}
We discuss the effect of the local field polarization in metallic
nanostructures where large field enhancements can be induced by
excitation of localized surface plasmons. To demonstrate the
importance of these effects, we study a few model systems in 2D
and 3D. As an experimental probe of this effect, we suggest to use
depolarization ratios in Surface Enhanced Raman Spectroscopy
(SERS) measurements. We attempt to relate the polarization of the
local field to the depolarization ratios in SERS conditions. This
approach could also lead to some better understanding of the SERS
mechanisms at a microscopic level.
\end{abstract}

\pacs{78.67.-n, 78.20.Bh, 78.67.Bf, 73.20.Mf}

\maketitle

\section{Introduction}
Raman depolarization ratios of molecular vibrations are one of the
most fundamental properties of inelastic light scattering in
liquids\cite{Loudon}; they are widely used in conjunction with
other types of spectroscopies to infer indirectly the symmetry of
molecular vibrational modes. Essentially, every vibration in a
molecule can be assigned to a given irreducible representation (or
linear combinations of them) of either the overall symmetry point
group of the molecule or one of its sub-structures. Once the
symmetry is known the Raman tensor of the vibration with respect
to a fixed molecular coordinates system is also known. A
collection of randomly oriented molecules will produce average
intensities for parallel $(I_{\parallel})$ or perpendicularly
$(I_{\perp})$ polarized scattering configurations in Raman
spectroscopy, from where a well defined depolarization ratio
$\rho\equiv I_{\perp}/I_{\parallel}$ follows. Both
$(I_{\parallel})$ and $(I_{\perp})$ only depend on tensor
invariants which do not change under rotations of the reference
frame. This is a canonical textbook problem in non-resonant
inelastic light scattering of liquids and poly-crystalline
solids\cite{Loudon}, which has also its exact counterpart in
elastic (Rayleigh) scattering (with the linear polarizability
tensor playing the role of the Raman tensor)\cite{Long}. There is
ample theoretical and experimental understanding of this effect in
Raman scattering, which started with the pioneering work of Porto,
Skinner, and Nielsen\cite{Porto,Skinner} in the 60's in the
angular dependence of inelastic light scattering of simple
liquids.

Notwithstanding, under Surface Enhanced Raman Scattering (SERS)
conditions in liquids (colloids) depolarization ratios of Raman
signals of the analyte molecules (typically dyes) are most of the
time anomalous and, in addition, show some laser wavelength
dependence. It is interesting to understand the origin of these
anomalies for such a fundamental physical property of the
scattering process like the depolarization ratio. The problem is
intimately related to the issue of {\it local fields} in SERS and
in plasmon-supporting structures in general. SERS is made possible
by highly localized plasmon resonance excitations called
hot-spots, which achieve large optical amplifications and boost
Raman scattering signals by several orders of
magnitude\cite{Moskovits,Otto}. Highly (sub-wavelength) localized
hot-spots have been proposed and studied both
theoretically\cite{shalaev,shalaev2} and
experimentally\cite{eric1,Mos1,Mos2}. It is the purpose of this
paper to shed some light on the polarization aspects of hot-spots
and how they affect basic properties of the symmetry of the
scattering. We also explore how these effects could be used in
principle to learn more about the nature of hot-spots, local
fields, and laser-field amplifications by plasmons.

We would like to address in the paper two main issues:
\begin{itemize}
\item Local-field polarization effects are often overlooked in
most treatments of the SERS effect, but should be important for
many essential properties of the scattering process. We shall show
that depolarization ratios are mainly determined by plasmons in a
SERS environment. The most striking demonstration of this effect
is the breakdown of normal depolarization ratios in single
isolated objects, as demonstrated in the next sections. \item
Depolarization ratios can, in principle, assert something about
the SERS enhancement mechanism itself and, in addition, the
equally important issue of molecular orientation and geometry of
the analyte on the metallic surface.
\end{itemize}

All these issues will be addressed with a variety of examples for
different geometries and assumptions, together with some
experimental evidence for anomalous depolarization ratios at the
end.

\section{Basic definitions}

SERS enhancements by one or several objects depend on the geometry
of the objects themselves, their symmetry, and their orientation
in space with respect to the fixed scattering polarization
directions in the laboratory frame. In addition, the scattering
can come from a single analyte (molecule) in a specific place on
the surface of the metal, or from the multiple contributions of a
distribution of molecules in many places. Last, but not least, the
Raman tensor of the analyte can be of many different types
(symmetries); hence the orientation of the main axes of the tensor
with respect to the metal surface at specific points has to be
specified also. In effect, the possibilities are endless. We need
to start by imposing some restrictions on the type of objects,
Raman tensors, and molecular orientations at the surface that we
will examine.

We first address the issue of the origin of the SERS signal in
terms of single/multiple dyes contributions. Interpretations of
the data based on single molecule detection have been claimed many
times in the past\cite{Kneipp_review,Kneipp,Nie}. If single
molecule detection is achievable then the polarization properties
of the local field at the position of the molecule will be the
only parameter that counts. The same situation happens if the
presence of analyte is widespread over the metal but there are
high intensity hot-spots dominating the signal. In such a
situation, the polarization properties of a specific place in
space and/or the orientation of the Raman tensor of one or a few
molecules are the main factors determining the symmetry of the
scattering event. The polarization properties would then be an
average (over time or space) of many of these hot-spots.
Polarization studies in SERS in single molecule limits are very
rare\cite{Brus,Brus2} and not fully understood in our opinion.

The most common experimental situation, however, is one in which
there are many contributions from many analytes which are
spatially distributed in different places and, accordingly, are
exposed to different local fields. This is for example the case of
colloidal solutions of gold or silver particles at relatively high
$(\sim 100$ nM) concentration of analyte and measured in
immersion. If conditions are chosen so that the solution is
stable, we can rule out the presence of large clusters of
particles (which would precipitate) and the signal must therefore
come from single particles or possibly pairs of particles (either
stable pairs or dynamical pairs formed during close encounters of
two particles). In order to observe a good SERS signal from such
solutions, higher analyte concentrations are required,
corresponding typically to $\sim$10$^2$-10$^4$ molecules per
particle. To study the polarization effects in such systems, one
has to consider the signals of all these molecules and also
average over all possible orientations of the particles. This is
only possible in simple cases and it is the main reason for the
restrictions we shall impose in the modelling.

\subsection{Basic geometries and Raman tensors}
We shall discuss here a few model cases: spherical particles,
elliptical particles, and dimers (two particles in close
proximity). In each case, we will model the SERS depolarization
ratio, which can be measured experimentally, under different
assumptions.

There are three main factors that can affect the calculated
depolarization ratio $\rho$.

\begin{itemize}
\item The Raman tensor of the probe molecules: We will consider
the cases of an isotropic Raman tensor ($Iso$) and of a uniaxial
tensor. These are two extreme cases of a variety of molecular
symmetries present in real analytes. Many common SERS analytes
(like rhodamine 6G) have strongly uniaxial Raman tensors for most
of their vibrations.

\item The adsorption configuration of the molecule: This is
obviously an important factor when the Raman tensor is not
isotropic. For the uniaxial case, we therefore need a further
assumption about the way the molecule adsorbs on the metal surface
and, in particular, how the main axis is oriented with respect to
the surface. We will consider the two extreme cases of a molecule
adsorbed with its axis tangential ($Tan$) or normal ($Norm$) to
the surface. We will also consider the case where the adsorption
geometry is random ($Rand$). In this latter case, all possible
orientations of the molecules are considered and averaged.

\item Finally, the way the molecule couples to the local field
polarization: This is also a very important aspect because it is
directly related to the SERS enhancements mechanism and has in our
opinion been overlooked.
\end{itemize}

We shall have a small digression on this latter issue here. If we
consider a molecule with a highly uniaxial Raman tensor oriented
along $\vec{e}_m$, then the excited dipole has the form (case
$A$):
\begin{equation}
\vec{p}^{A}=\alpha \left( \vec{E} \cdot \vec{e}_m \right)
\vec{e}_m, \label{casea}
\end{equation}
where $\vec{E}$ is the local electric field and $\alpha$ is the
main component of the uniaxial polarizability. This is a case
often found in normal Raman scattering. The intensity scattered
and detected in the far field in a given direction $\vec{e}_d$,
along a given polarization $\vec{e}_p$ $(\vec{e}_p \perp
\vec{e}_d)$ is then $I\propto
\left|\vec{p}\cdot\vec{e}_p\right|^2$. However, if we use this
expression in SERS conditions, the SERS signal is only
proportional to the power of two of the local field enhancement
factor (and not power of four as usually assumed\cite{Moskovits}).
In order to obtain a power of four in the scattering efficiency,
we need to model the enhancement process due to re-emission of
this dipole. One could use the following expression (case $B$):
\begin{equation}
\vec{p}^{B}= \frac{\alpha |\vec{E}|}{|\vec{E}_0|} \left( \vec{E}
\cdot \vec{e}_m \right) \vec{e}_m.\label{caseb}
\end{equation}
Anywise, this implies that the enhancement in re-emission has no
polarization dependence. Another view is to assume that the
excited dipole couples only to the local field enhancement along
its direction; we then get (case $C$):
\begin{equation}
\vec{p}^{C} = \frac{\alpha}{|\vec{E}_0|} \left( \vec{E} \cdot
\vec{e}_m \right)^2 \vec{e}_m.\label{casec}
\end{equation}
In both cases, we have also assumed that the local field
enhancements relevant to re-emission were the same as those
derived for excitation. This is a common assumption but is not
necessarily valid. Assuming we know the local field enhancements
for a given geometry, we can then calculate the polarization
property of the SERS signal for each of the cases $A$, $B$, and
$C$ at each point of the surface $S$ of our metallic structure/s.
These can then be integrated over the surface, assuming a
continuous distribution of probe molecules of type $Iso$, $Tan$,
$Norm$, or $Rand$.

\subsection{Symmetries and averages}

As an example, and to provide a basis for further discussion, we
carried out calculations for simple model systems in 2 dimensions
(2D). The structures are in the plane ($yOz$) and the exciting
beam is coming from the perpendicular direction ($Ox$), with a
field polarization along $z$. The local fields at the surface are
calculated within the electrostatics approximation. The
calculations were carried out numerically using finite element
modelling described elsewhere\cite{FEMLAB}. We assume that the
signal is detected in the far field along ($Ox$) and analyzed for
parallel ($Oz$) and perpendicular ($Oy$) polarizations. For a
given structure, we calculated $I_{\parallel}$ and $I_{\perp}$ for
each of our assumptions. $\vec{n}$ ($\vec{t}$) denotes the unit
normal (tangential) vector at any point on the metallic surface.
We therefore used the following expressions:
\begin{equation}
I_{\parallel}^{Iso-A}=\int_S \left| E_z \right| ^2 dS,
\end{equation}
\begin{equation}
I_{\parallel}^{Iso-B}=I_{\parallel}^{Iso-C}=\int_S \left|
\vec{E}\right| ^2 \left| E_z \right| ^2 dS,
\end{equation}
\begin{equation}
I_{\parallel}^{Norm-A}=\int_S \left| \vec{E}\cdot\vec{n}\right|^2
n_z^2 dS,
\end{equation}
\begin{equation}
I_{\parallel}^{Norm-B}=\int_S \left| \vec{E}\right|^2 \left|
\vec{E}\cdot\vec{n}\right|^2 n_z^2 dS,
\end{equation}
and
\begin{equation}
I_{\parallel}^{Norm-C}=\int_S \left| \vec{E}\cdot\vec{n}\right|^4
n_z^2 dS.
\end{equation}
Similar expressions for $Tan-A, -B, -C$ replacing $\vec{n}$ with
$\vec{t}$ are also obtained. Expressions for $I_{\perp}$ are
obtained by replacing $z$ subscripts with $y$. We will also for
each case estimate the total SERS intensity by means of:
\begin{equation}
I_{Tot}=I_{\parallel}+I_{\perp}.
\end{equation}

For structures without rotational symmetry, the calculations were
repeated for 50 different orientations, parameterized by an angle
$\beta$. The total signals are then obtained by summing over all
possible orientations. In our structures, we need only to consider
$\beta$ from 0 to $\pi/2$ by symmetry, so we have:
\begin{equation}
I_{\parallel}=\int_0^{\pi/2} I_{\parallel}(\beta) d\beta,
\end{equation}
and the depolarization ratios is then calculated according to
\begin{equation}
\rho=\frac{I_{\perp}}{I_{\parallel}}.
\end{equation}

In most simulations of metallic structures presented in the
literature, such as for ellipses or dimers, only one or two
possible orientations are considered. In liquids, contributions
from all orientations are averaged and the resulting intensity
profile can be very different to that of a single configuration.
Independent of polarization effects, we believe that the
orientation-averaged SERS intensity profiles presented here
provides a new insight into plasmon resonances in liquids.

For the $Rand$ cases, one has to consider all possible molecular
orientations. In 2D we restrict ourselves to:
\begin{equation}
\vec{e}_m=\cos(\gamma)\vec{e}_y+\sin(\gamma)\vec{e}_z.
\end{equation}
Estimating the intensities and averaging over $\gamma$, one finds
that the results $Rand-B$ and $Rand-C$ can be directly obtained
from that of $Iso-B$ (also equal to $Iso-C$), namely:
\begin{equation}
 \rho^{Rand-B}=\frac{3\rho^{Iso-B}+1}{\rho^{Iso-B}+3},
\label{aa}
\end{equation}
\begin{equation}
I_{Tot}^{Rand-B}=\frac{1}{2}I_{Tot}^{Iso-B},
\end{equation}
\begin{equation}
 \rho^{Rand-C}=\frac{5\rho^{Iso-B}+1}{\rho^{Iso-B}+5},
\label{bb}
\end{equation}
and
\begin{equation}
I_{Tot}^{Rand-C}=\frac{3}{8}I_{Tot}^{Iso-B}.
\end{equation}

As a reminder, in non-SERS conditions in 2 dimensions, $\rho$
should be 0 for an isotropic molecule and $1/3$ for the uniaxial
case ($Rand$).

For all calculations, we used the bulk dielectric constant of Ag
(a typical metal for SERS applications) as reproduced by a Drude
model with the best fit to tabulated data in the region of
interest\cite{Palik}, namely:

\begin{equation}
\epsilon(\lambda)= \epsilon_{\infty} - \frac{1}
{\lambda_{p}^2~\left(\frac{1}{\lambda^2}+\frac{i}{\Gamma\lambda}\right)},
\label{Drude}
\end{equation}
where $\epsilon_{\infty}=4$, $\lambda_{p}=141$ nm, $\Gamma=17000$
nm, and $\lambda$ is the wavelength. The surrounding medium is
water with $\epsilon_m=1.77$.

\section{The simplest case: SERS by a single symmetric object}
A simple object like a disk in 2D or a sphere in 3D has a well
defined main plasmon resonance. If the size of the object $(a)$ is
much smaller than the wavelength $(a\ll\lambda)$ of the light,
then the electrostatic approximation is valid and the resonance
occurs at the frequency where Re$(\epsilon(\lambda))\approx
-\epsilon_m$, or $-2\epsilon_m$ for a disk in 2D or a sphere in
3D, respectively. If $(a\sim\lambda)$ on the contrary, retardation
effects and higher order multipoles play an increasing role and
define additional resonances. The current understanding of the
SERS effect is that it benefits from red-shifted plasmon
resonances\cite{Us1} which are either shape/size-related or, more
often than not, coming from plasmon-plasmon interactions among
particles\cite{Xu,Calander}. These latter resonances are believed
to be the dominant contribution to the SERS signal in general.

The case of a 2D sphere (or disk) of radius $a$, is particularly
easy because of the rotational symmetry (no need for different
$\beta$'s in the averaging). It can also be easily solved
analytically in the electrostatic approximation. We consider an
Ag-disk with complex dielectric function $\epsilon(\lambda)$ in
water ($\epsilon_m=1.77$). The important parameter in such problem
is the relative dielectric constant $\epsilon_r=\epsilon /
\epsilon_m$. For an incident field $\vec{E}_0=1 \vec{e}_z$, the
local field at a point on the surface with coordinates $a, \theta$
is simply the superposition of the incident field and of an
induced dipolar field:
\begin{equation}
E_y(\theta)=k \sin(2\theta),
\end{equation}
\begin{equation}
E_z(\theta)=1-k \cos(2\theta),
\end{equation}
with
\begin{equation}
k=\frac{\epsilon_r -1}{\epsilon_r+1}.
\end{equation}
We can therefore derive the following analytical expressions:
\begin{equation}
\rho^{Iso-A}=\frac{|k|^2/2}{1+|k|^2/2},
\end{equation}
\begin{equation}
\rho^{Iso-B}=\frac{|k|^4/2+|k|^2/2}{1+|k|^4/2+3|k|^2/2+2
\text{Re}(k)^2},
\end{equation}
\begin{equation}
\rho^{Norm-A}=\rho^{Tan-A}=1/3,
\end{equation}
\begin{equation}
\rho^{Norm-C}=\rho^{Tan-C}=1/5,
\end{equation}
\begin{equation}
\rho^{Norm-B}=\frac{|\epsilon_r|^2+1}{5|\epsilon_r|^2+1},
\end{equation}
and
\begin{equation}
\rho^{Tan-B}=\frac{|\epsilon_r|^2+1}{|\epsilon_r|^2+5},
\end{equation}

where $\rho^{Rand-B}$ and $\rho^{Rand-C}$ can also be evaluated
analytically from the expression of $\rho^{Iso-B}$ using Eqs.
(\ref{aa}) and (\ref{bb}).

\begin{figure}[h]
 \centering{
  \includegraphics[height = 19.2 cm]{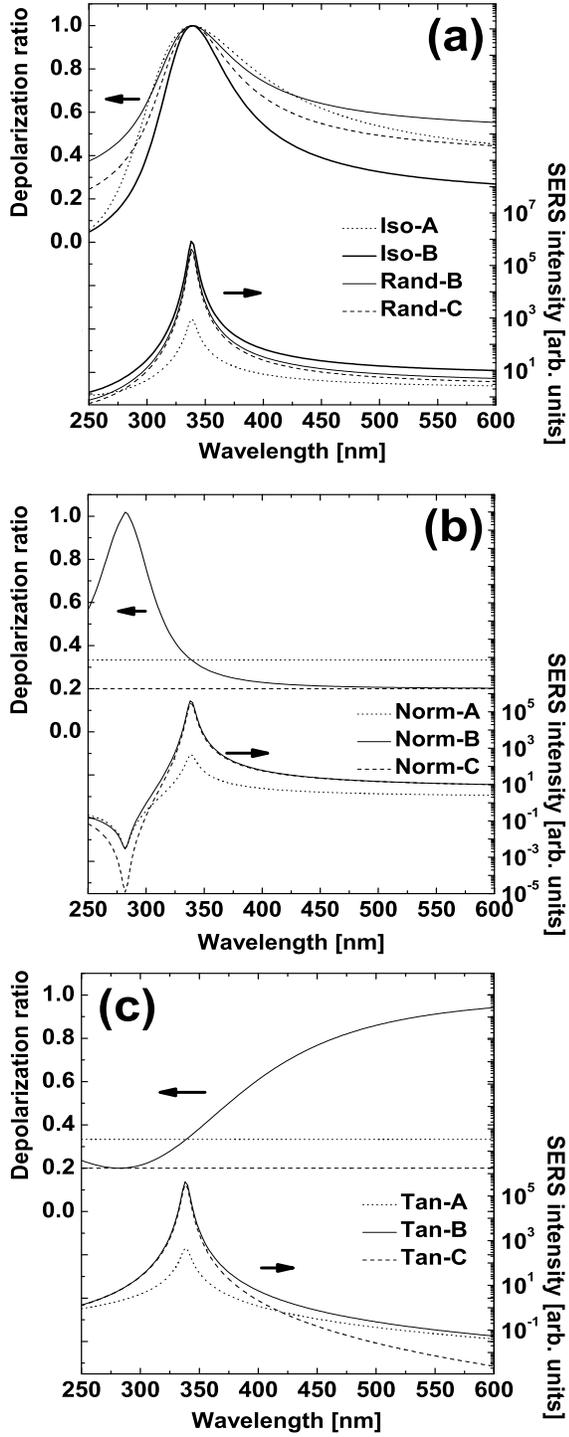}
  }
 \caption{Depolarization ratios and total SERS intensities
calculated for a 2D disk from different assumptions as a function
of wavelength. $Tan$, $Norm$, $Iso$ and $Rand$ define the type of
Raman tensor under consideration, as specified in the text (Sec.
II), while $A$, $B$, and $C$ account for the different models of
the scattering process defined in Eqs. (\ref{casea}),
(\ref{caseb}), and (\ref{casec}). Note that the SERS enhancements
are plotted on a log-scale.} \label{fig1}
\end{figure}

Figure \ref{fig1} shows the wavelength dependence of the
depolarization ratio under the various assumptions considered
here. An important aspect to help distinguish these cases is also
the total SERS intensity, which is not reflected in the values of
$\rho$. We know that the SERS signal is usually strong enough to
be detected, and any valid assumption should therefore predict a
reasonably strong SERS signal in addition to a correct
depolarization ratio. We show in Fig. \ref{fig1} the calculated
SERS intensity as a function of wavelength. As can be expected,
the three cases where the SERS enhancement is only proportional to
the square of the field amplitude (cases $A$) show a relatively
small SERS intensity. These three cases will therefore not be
studied any further. All the other cases show a strong enhancement
at the localized surface plasmon resonance of the object
($\text{Re}(\epsilon)=-\epsilon_m$, $\lambda\sim 339$ nm in this
case). For $Tan-B$ and $Tan-C$, this enhancement drops sharply
beyond this resonance, while a small enhancement still exists in
the other cases.

This is a very interesting case for more than one reason. It is
interesting to see how a large variety of cases is observed for
$\rho$ in an object which is arguably one of the simplest SERS
enhancing object we can think of. In some cases, $\rho=1/3$ and is
independent of wavelength ($Norm-A$, $Tan-A$) or 1/5 ($Norm-C$,
$Tan-C$). In other cases, it increases to 1 close to the surface
plasmon resonance and decreases slowly beyond that, towards a
limiting value of $1/5$ ($Iso-B$), $5/13$ ($Rand-C$), or $1/2$
($Rand-B$). $Tan-B$ and $Norm-B$ show a somewhat different
behavior, the first showing an increase towards 1, while the
latter peaks sharply at 1 before the surface plasmon resonance.

Conceptually, it is interesting to see that for the case of an
isotropic probe (which should exhibit $\rho=0$ in free space)
$\rho$ can be as large as 1 as a result of the interaction with
the metallic disk ($Iso-B$ at resonance). This is surprising
because the disk itself is also isotropic, but {\it the anisotropy
is introduced by excitation of non-isotropic dipolar plasmon
resonances in the object. This demonstrates clearly that the
localized plasmon resonances are crucial in the polarization
mechanisms and cannot be excluded from the discussion}. It is even
clear from Fig.\ \ref{fig1} that the plasmon resonances, and not
the probe molecules, are the dominant factor in the origin of the
depolarization ratios; what counts is {\it the symmetry of the
plasmon resonance for the polarization properties of the
scattering}. This summarizes in many ways one of the most
important points of this paper.

\begin{figure}[h]
 \centering{
  \includegraphics[height = 16 cm]{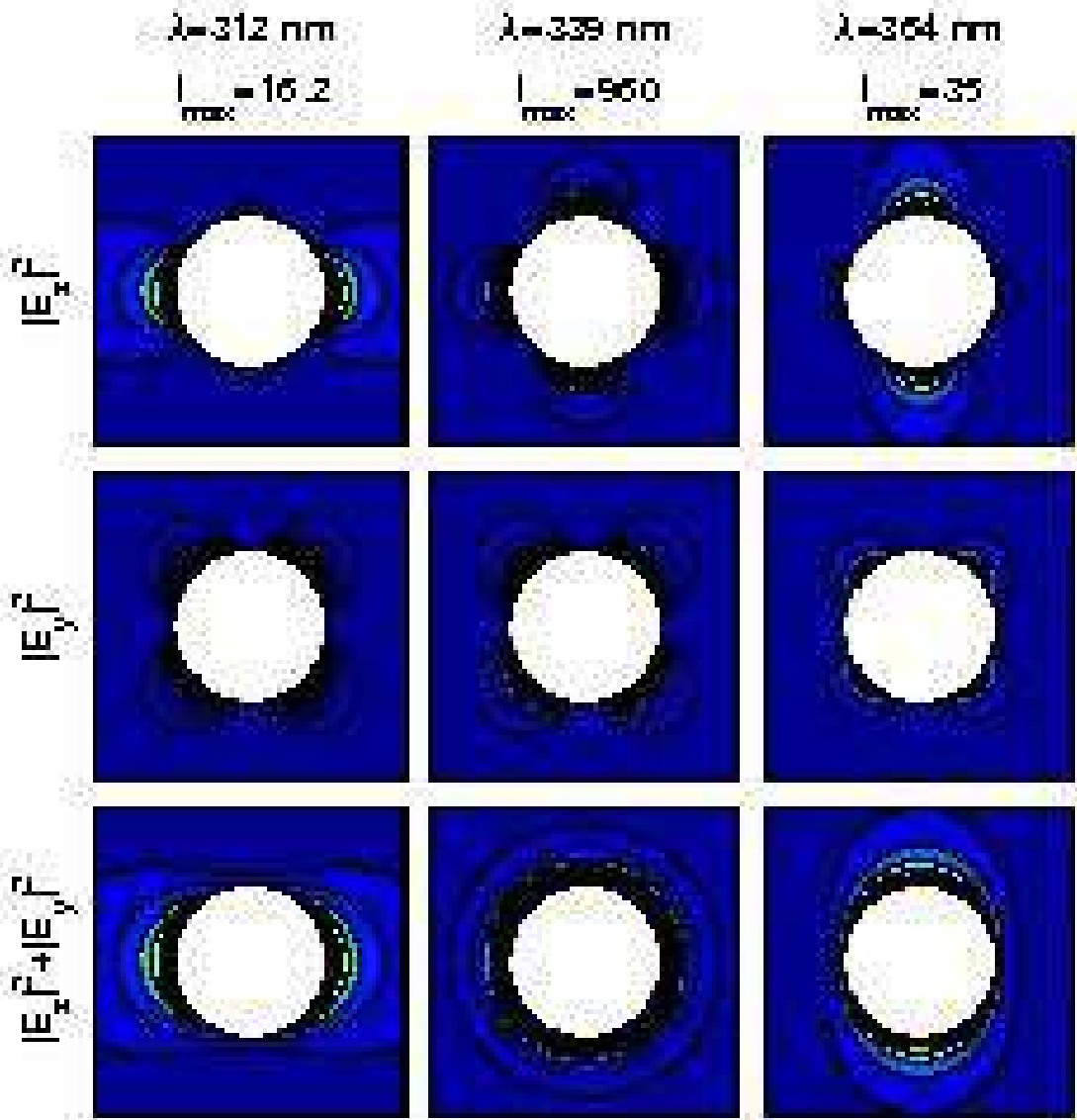}
 }
\caption{Intensity patterns in a 2D disk at three different
wavelengths (columns) for the field component along $z$ (first
row), $y$ (second row), and for the total intensity (third row).
The excitation polarization of the external field is along $z$.
For a scatterer with an isotropic Raman tensor, for example, the
integrated intensity on the surface of the disk for $|E_y|^2$
divided by that for $|E_z|^2$ defines the depolarization ratio
$\rho$. By visual inspection, it is evident that the
depolarization ratio will be smaller at 312 and 364 nm than at 339
nm, where the disk has its main resonance. Colors mean different
intensities for the different columns; we provide the maximum
field intensity for each wavelength at the top of each column. Red
(blue) means high (low) intensity in a linear color scale. See the
text for further details.} \label{fig2}
\end{figure}

It is interesting to see, in addition, the reasons for the
influence of plasmon resonances in terms of field patterns for the
simplest possible case treated in this section. In a way, all the
other effects are further degrees of sophistication of this simple
example. Figure \ref{fig2} shows for this 2D case the intensity
patterns on the surface for an incident polarization along $z$ of
$\left|E_z\right|^2$, $\left|E_y\right|^2$, and the total field
$\left|\vec{E}\right|^2$, at three different wavelengths chosen to
be below, at, and above the main surface plasmon resonance of the
object $(\lambda=339$ nm). The case can be solved analytically and
all the patterns can be interpreted as a direct superposition of
the external field and an equivalent induced dipole on the disk.
It is the complex polarizability of the metal that makes the
dipole response non-trivial because of the relative amplitude and
phase shift of the response with respect to the exciting field.
From Fig. \ref{fig2} we can see that {\it even if we had molecules
with an isotropic Raman tensor (emitting in the same direction as
the excitation), the depolarization ratio will be changing with
wavelength}. Ignoring the Stokes shift, the integrated intensity
on the surface for $\left|E_y\right|^2$ divided by that for
$\left|E_z\right|^2$ will give the depolarization ratio at that
wavelength. It is easy to infer by visual inspection that the
depolarization ratio will be small at $\lambda=312$ nm, large at
resonance $(\lambda=339$ nm), and small again at longer
wavelengths $(\lambda= 364$ nm). It is large (close to 1) at
resonance because the induced dipolar response completely
dominates the signal. The (analytical) example in Fig. \ref{fig2}
is possibly the simplest example of plasmons affecting the Raman
depolarization ratio by changing the nature of the coupling
between the laser and the molecules through the local field. In
practise the Raman tensor of the molecule itself will play a role,
adding an additional degree of complexity to the problem.

As we shall show later for a particular analyte with a uniaxial
Raman tensor, SERS experiments on colloidal solutions show that
the depolarization ratio indeed changes with wavelength and ranges
from 0.4 to 0.7 in our measurements. This observation therefore
rules out some of the cases studied here. We can identify three
cases which are compatible with this observation and predict a
reasonable SERS intensity: $Iso-B$, $Rand-B$, and $Rand-C$. In
order to determine which of these cases is the most realistic, one
first needs to look at more realistic models. It is in particular
important to determine the effect on $\rho$ of different particle
shapes and to assess the validity of using a 2D model to infer 3D
properties. Finally retardation effects (not taken into account in
the electrostatics approximation) can have some influence on the
plasmon resonances and therefore also on $\rho$. We address some
of these issues in the next sections.

\section{Shape effects}

\begin{figure}[h]
 \centering{
  \includegraphics[height = 16 cm]{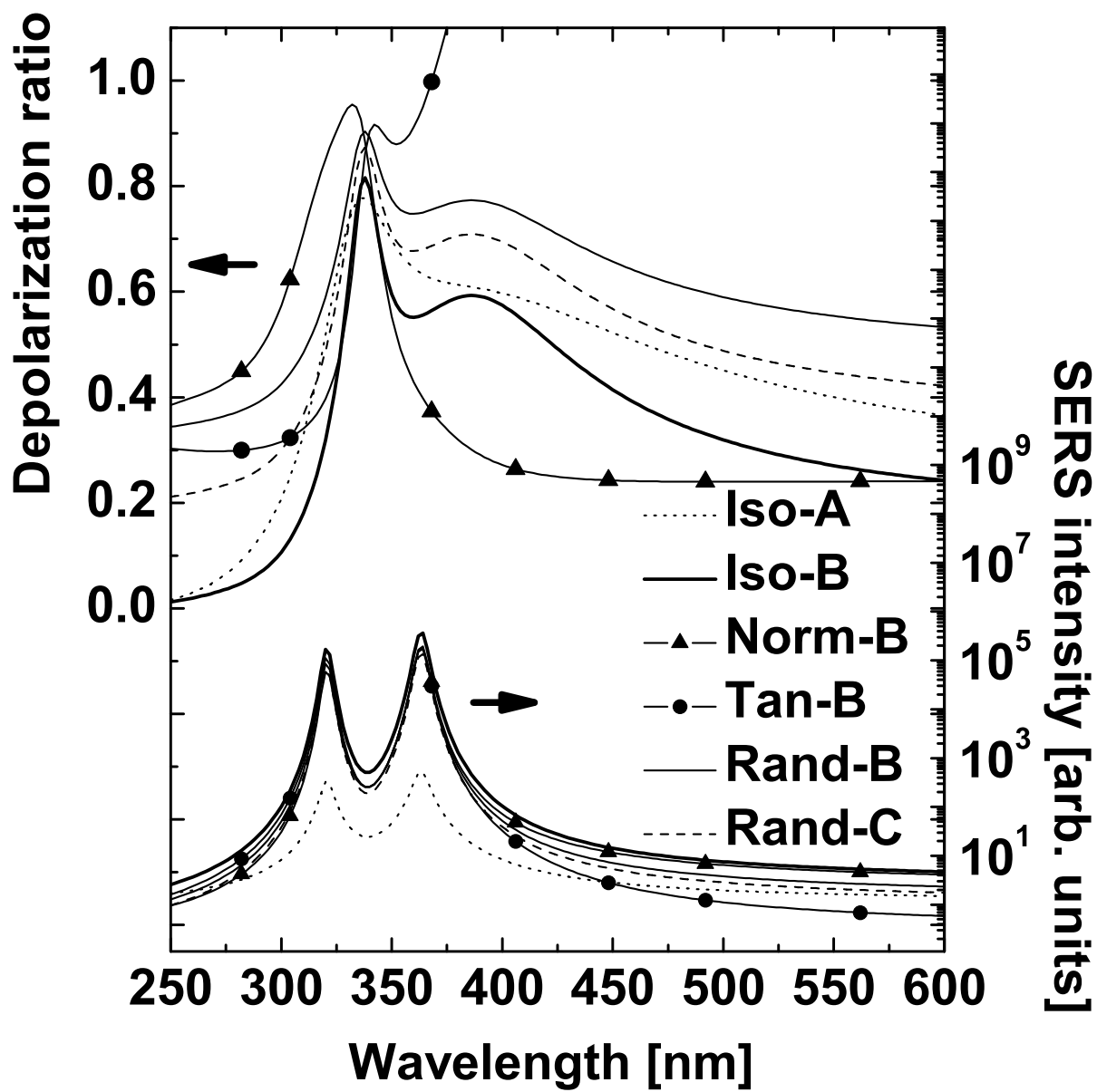}
 }
\caption{Depolarization ratios (top) and total SERS intensities
(bottom) as a function of wavelength calculated for an ellipse for
some of the models presented in Fig. \ref{fig1}. As opposed to the
case of a disk in Fig. \ref{fig1}, an ellipse has two main
shape-related plasmon resonances in the SERS intensity profile.
The $Tan-B$ model for the orientation of the molecule and the
scattering mechanism gives depolarization ratios which go well
above 1 across the visible range.}
 \label{fig3}
\end{figure}

To understand at least qualitatively the effects of shape, we will
consider the case of an ellipse with an aspect ratio between long
and short axes of 1.5. Because we now loose the rotational
symmetry, we consider the contributions to the SERS signal from
ellipses with 50 possible orientations. Figure \ref{fig3} shows
the calculated total SERS intensity under different assumptions.
As expected, the plasmon resonance is now split into two
resonances: one red-shifted at 363~nm, the other blue-shifted at
320~nm. In many instances in the literature, excitations along one
or the other principal axis are usually considered separately and
only one resonance is observed at a time. Both resonances are
observed here because all possible orientations are considered and
summed. Figure \ref{fig3} also shows the predicted depolarization
ratios. Again, a wide variety of different behaviors are observed,
but we will concentrate on the three cases identified previously:
$Iso-B$, $Rand-B$, and $Rand-C$. All three show a similar pattern
with $\rho$ peaking for $\lambda=340$~nm to a value at $0.8-0.9$,
then at $\lambda=388$~nm to a value of $0.6-0.8$, and then
decreasing to the same limiting value as that observed for the
disk.

{\it It is interesting to note that the peaks of $\rho$ do not
correlate exactly with the plasmon resonances observed in the SERS
intensity. The first peak remains at the plasmon resonance of a
sphere, while the second appears at even longer wavelength than
the red-shifted plasmon resonance. Also, the maximum value of
$\rho$ is now slightly smaller than 1.} This can have some
important consequences for colloidal solutions with a distribution
of sizes and shapes. For a given excitation wavelength
$\lambda_L$, the SERS signal should have a larger contribution
from particles in resonance with this wavelength. If there are
many such particles, the depolarization ratio will be determined
by these and will then correspond to the dip in Fig.\ \ref{fig3}
at 363~nm (i.e. $\rho$ is smaller than its maximum value).
However, if $\lambda_L$ is longer than most particle resonances,
the measured $\rho$ should correspond to values to the right of
the dip in Fig.\ \ref{fig3}. This could lead to an increase in
$\rho$ if $\lambda_L$ remains close to most resonances, or to a
decrease at much longer wavelength. Such considerations are
important when interpreting experimental results in colloidal
solutions because polydiversity is inevitable.

\section{Case of a 3D sphere}

It could be argued that conclusions gained from 2D models on
depolarization ratios should be taken with care; it is necessary
to estimate the changes that will arise from looking at real 3D
cases to gain some insight on how reliable the qualitative
predictions of 2D models are. Numerical simulations in 3D are much
more computationally demanding than the 2D cases. To assess the
importance of the corrections arising from a full 3D model, we
will therefore study the case of a 3D sphere and compare
qualitatively the results to the 2D case. Similar to the 2D disk,
the 3D sphere problem can be solved analytically in the
electrostatics approximation. For an incident field
$\vec{E}_0=1\vec{e}_z$, the local field outside the sphere is
simply the superposition of the incident field and of an induced
dipolar field:
\begin{equation}
\vec{E}=3\kappa\cos(\theta) \vec{e}_r +(1-\kappa)\vec{e}_z,
\end{equation}
where $(r,\theta,\phi)$ are the standard spherical coordinates and
\begin{equation}
\kappa=\frac{\epsilon_r-1}{\epsilon_r+2}.
\end{equation}
Calculations can be carried out in a similar way as for the 2D
case. To illustrate this, we describe below the $Norm-C$, $Tan-C$,
and $Iso-B$ cases.

In the $Norm-C$ case, the molecules are assumed to be adsorbed
with their main axis normal to the metal surface. The parallel and
perpendicular intensities are therefore given by:
\begin{equation}
I_{\parallel}^{Norm-C}=\int_S |\vec{E} \cdot \vec{n} |^4 (\vec{n}
\cdot \vec{e}_z )^2 dS=\frac{2}{7}\left| 1+2\kappa \right|^4,
\end{equation}
and
\begin{equation}
I_{\perp}^{Norm-C}=\int_S |\vec{E} \cdot \vec{n} |^4 (\vec{n}
\cdot \vec{e}_y )^2 dS=\frac{2}{35}\left| 1+2\kappa \right|^4.
\end{equation}
The depolarization ratio is then independent of wavelength and
equal to $\rho^{Norm-C}=1/5$, which is the same result as that
obtained in 2D.

For the case of $Tan-C$, we assume the molecule is adsorbed with
its axis tangential to the metallic surface. However, in 3D, this
does not define completely the orientation of the axis, and we
therefore need to average over all possible orientations of the
molecular axis (tangential to the surface). For this, we assume
the axis to be along the unit vector:
\begin{equation}
\vec{e}_m=\cos \gamma \vec{e}_\theta + \sin \gamma \vec{e}_\phi.
\end{equation}
We then calculate:
\begin{eqnarray}
I_{\parallel}^{Tan-C}(\gamma)&=&\int_S |\vec{E} \cdot \vec{e}_m
|^4 (\vec{e}_m \cdot \vec{e}_z )^2 dS\\
&=&\frac{32}{35} \cos^6 \gamma \left| 1-\kappa \right|^4,
\end{eqnarray}
and
\begin{eqnarray}
I_{\perp}^{Tan-C}(\gamma)&=&\int_S |\vec{E} \cdot \vec{e}_m |^4
(\vec{e}_m \cdot \vec{e}_y )^2 dS\\
&=&\frac{8}{105} \left| 1-\kappa \right|^4 \cos^4 \gamma
(1+6\sin^2 \gamma).
\end{eqnarray}

Averaging over $\gamma$ ($0 \le \gamma \le  2\pi$), we obtain:
\begin{equation}
I_{\parallel}^{Tan-C}=\frac{2}{7}\left| 1-\kappa \right|^4,
\end{equation}
and
\begin{equation}
I_{\perp}^{Tan-C}=\frac{2}{35}\left| 1-\kappa \right|^4.
\end{equation}
The depolarization ratio is therefore again $\rho^{Tan-C}=1/5$,
independent of wavelength and equal to the 2D case.

In the 2D section, we identified the cases $Iso-B$, $Rand-B$, and
$Rand-C$ as the most interesting. The $Iso-B$ case can be
calculated analytically, to wit:
\begin{eqnarray}
I_{\parallel}^{Iso-B}&=&\frac{2}{105}|1-\kappa|^4 \\&~&\left[
(3|\epsilon_r|^2+4)(5|\epsilon_r|^2-2+4\text{Re}(\epsilon_r) )+56
\right]\nonumber,
\end{eqnarray}
and
\begin{equation}
I_{\perp}^{Iso-B}=\frac{6}{35}| 1-\kappa |^2 |\kappa|^2
(3|\epsilon_r|^2+4).
\end{equation}
Hence, the ratio:
\begin{equation}
\rho^{Iso-B}=\frac{|\epsilon_r-1|^2
(3|\epsilon_r|^2+4)}{(3|\epsilon_r|^2+4)
(5|\epsilon_r|^2-2+4\text{Re}(\epsilon_r) )+56}.
\end{equation}

For the $Rand-B$ and $Rand-C$ cases, one needs to consider
molecules with a random orientation of their axis:
\begin{equation}
\vec{e}_m=\sin \gamma \cos \delta \vec{e}_x+ \sin \gamma \sin
\delta \vec{e}_y +\cos \gamma \vec{e}_z,
\end{equation}
with $0<\gamma<\pi$ and $0<\delta<2\pi$. Averaging over these
parameters, one can show that:
\begin{equation}
\rho^{Rand-B}=\frac{1+4\rho^{Iso-B}}{3+2\rho^{Iso-B}},
\end{equation}
\begin{equation}
I_{Tot}^{Rand-B}=\frac{12}{15}I_{\bot}^{Iso-B}+\frac{8}{15}I_{\parallel}^{Iso-B},
\end{equation}
\begin{equation}
\rho^{Rand-C}=\frac{1+6\rho^{Iso-B}}{5+2\rho^{Iso-B}},
\end{equation}
and
\begin{equation}
I_{Tot}^{Rand-C}=\frac{12}{35}I_{\bot}^{Iso-B}+\frac{16}{15}I_{\parallel}^{Iso-B}.
\end{equation}

\begin{figure}[h]
 \centering{
  \includegraphics[height = 12 cm]{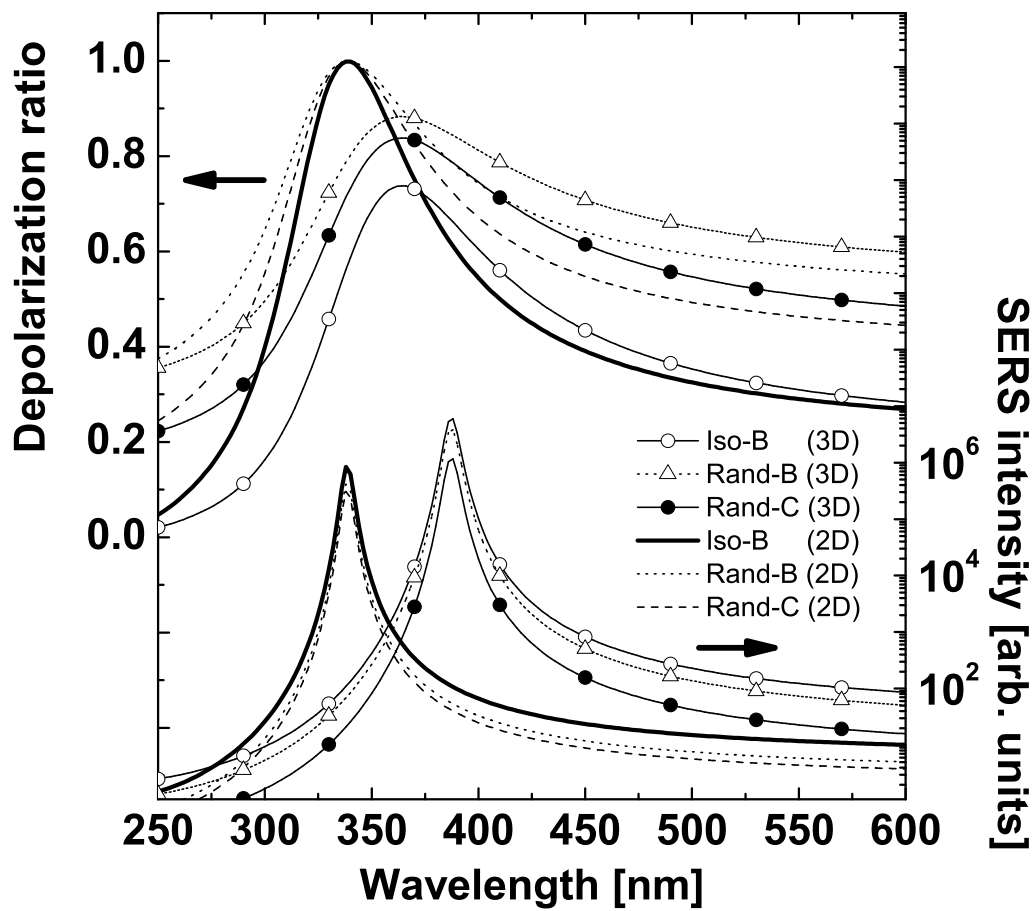}
 }
\caption{Comparison between the depolarization ratios (top) and
the total SERS intensities (bottom) calculated for a 3D sphere
with those of a 2D disk. The curves with symbols (lines) are the
3D (2D) cases for models $Iso-B$, $Rand-B$, and $Rand-C$,
respectively. The depolarization ratios together with the SERS
intensities in this figure complete the qualitative/quantitative
comparison of the effect of dimensionality on these quantities.}
 \label{fig4}
\end{figure}

These results are plotted in Fig. \ref{fig4}, with the equivalent
2D results for comparison. We first notice, as expected, that the
resonance profile of the SERS intensity is further red-shifted in
the 3D case, peaking at 388~nm (Re$(\epsilon_r)=-2$) instead of
339~nm (Re$(\epsilon_r)=-1$). Regarding the depolarization ratios,
they show the same qualitative behavior as in 2D with some small
quantitative changes: they are smaller for the 3D case below the
plasmon resonance, and slightly larger beyond this resonance
tending towards the same value at long wavelength. Also, the
maximum depolarization ratio is no longer 1 as in the 2D case but
around 0.7-0.9 depending on the case. Finally, the maximum of
$\rho$ now occurs slightly before the resonance at around 368~nm.
For the cases $Rand-B$ and $Rand-C$, $\rho$ remains quite high, at
least larger than 0.5 beyond the resonance, while it goes down to
around 0.3 for $Iso-B$.

The overall conclusion of translating the simulation from 2D to 3D
shapes is that the essential qualitative features are already
contained in the 2D simulation. The quantitative picture will, of
course, be different. But the core qualitative (and
semi-quantitative) effect of the main plasmon resonance is already
contained in the 2D simulation.

\section{Retardation effects}

So far, all the calculations have been performed within the
electrostatics approximation. This is strictly speaking valid for
very small objects $a \ll \lambda$, where the electric field can
be assumed to be constant over the object. In practise,
retardation effects can become non-negligible in the visible range
for objects of dimensions $\approx 30$~nm ($\lambda/10$).
Retardation effects tend to red-shift the plasmon resonances and
add more structure to the resonance profile. In order to appraise
the effect of retardation on the depolarization ratios, we used
Mie theory \cite{Mie08} to calculate exact results for the simple
case of 3D spheres. We follow the treatment of Mie theory given by
Bohren and Huffman \cite{Bohren83}. For a sphere of radius $a$, we
calculate the exact value of the field at the surface
$\vec{E}(a,\theta,\phi)$. We can then evaluate the relevant
integrals to calculate SERS intensities and depolarization ratios
as before. In particular in the $Iso-B$ case, we calculate:
\begin{eqnarray}
I_{\parallel}^{Iso-B}=\int_S |E|^2|E_z|^2 dS, \\
I_{\perp}^{Iso-B}=\int_S |E|^2|E_y|^2 dS, \\
I_{\perp 2}^{Iso-B}=\int_S |E|^2|E_x|^2 dS.
\end{eqnarray}
$I_{\bot 2}$ is a magnitude that cannot be measured when detecting
along the excitation direction $x$, but will be useful
nevertheless to calculate other ratios. The depolarization ratio
is then
\begin{equation}
\rho^{Iso-B}=\frac{I_{\perp}^{Iso-B}}{I_{\parallel}^{Iso-B}}.
\end{equation}
By considering molecules of random orientations, one can also show
that:
\begin{eqnarray}
I_{\parallel}^{Rand-B}=\frac{2}{15} \left[
3I_{\parallel}^{Iso-B}+I_{\perp}^{Iso-B}+I_{\perp 2}^{Iso-B}\right], \\
I_{\perp}^{Rand-B}=\frac{2}{15} \left[
I_{\parallel}^{Iso-B}+3I_{\perp}^{Iso-B}+I_{\perp 2}^{Iso-B}\right],\\
\end{eqnarray}
and
\begin{eqnarray}
I_{\parallel}^{Rand-C}=\frac{2}{35} \left[
5I_{\parallel}^{Iso-B}+I_{\perp}^{Iso-B}+I_{\perp 2}^{Iso-B}\right], \\
I_{\perp}^{Rand-C}=\frac{2}{35} \left[
I_{\parallel}^{Iso-B}+5I_{\perp}^{Iso-B}+I_{\perp 2}^{Iso-B}\right]. \\
\end{eqnarray}
The corresponding depolarization ratios $\rho^{Rand-B}$ and
$\rho^{Rand-C}$ can then be evaluated.

\begin{figure}[h]
\centering{
  \includegraphics[height = 12 cm]{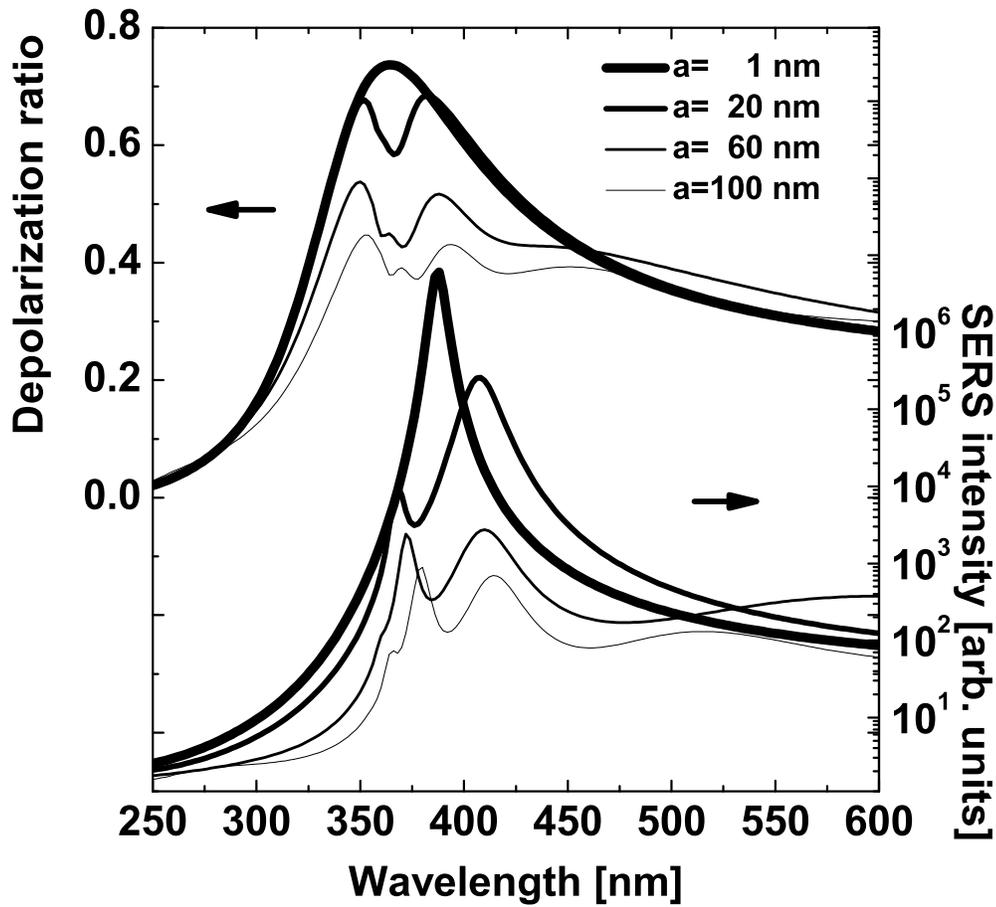}
 }
\caption{Depolarization ratio (top) and SERS intensities (bottom)
calculated using exact Mie theory for 3D spheres of various radii
$a$ from 1 to 60 nm. Thicker (thinner) lines represent smaller
(larger) radii. The results are shown for the $Iso-B$ case.}
 \label{fig5}
\end{figure}

In Fig.\ \ref{fig5} we display the results for the $Iso-B$ case,
for sphere radii in the range $a=1$ to $a=60$~nm. As expected, the
results for $a=1$~nm are identical to those obtained from the
electrostatics approximation. As $a$ increases, retardation
effects lead to a number of modifications: First the plasmon
resonance is split into several resonances, one of which
red-shifts more the larger the radius. The maximum SERS intensity
also decreases as $a$ increases. Concerning the depolarization
ratios, the most interesting point is that it shows virtually no
change beyond its original resonance at 368~nm. In particular, the
most red-shifted resonances of the intensity (which are the ones
most relevant to SERS) do not affect $\rho$. Instead, $\rho$ is
modified in the regions of the secondary resonances. A closer look
at the data indicates that $\rho$ shows a dip for each of the
resonances in the intensity (except the most red-shifted one). For
example, for $a=30$~nm, two dips at 358 and 374~nm are observed
for $\rho$, and they correspond to peaks in the intensity profile.
However, the most red-shifted resonance (at $\sim$432~nm) does not
affect the profile of $\rho$. We conclude that in the regions of
interest to SERS, retardation effects do not in general affect the
main conclusions on depolarization ratios gained from the simpler
electrostatic approach.

\section{Coupled resonances - Case of a dimer}

\begin{figure}[h]
 \centering{
  \includegraphics{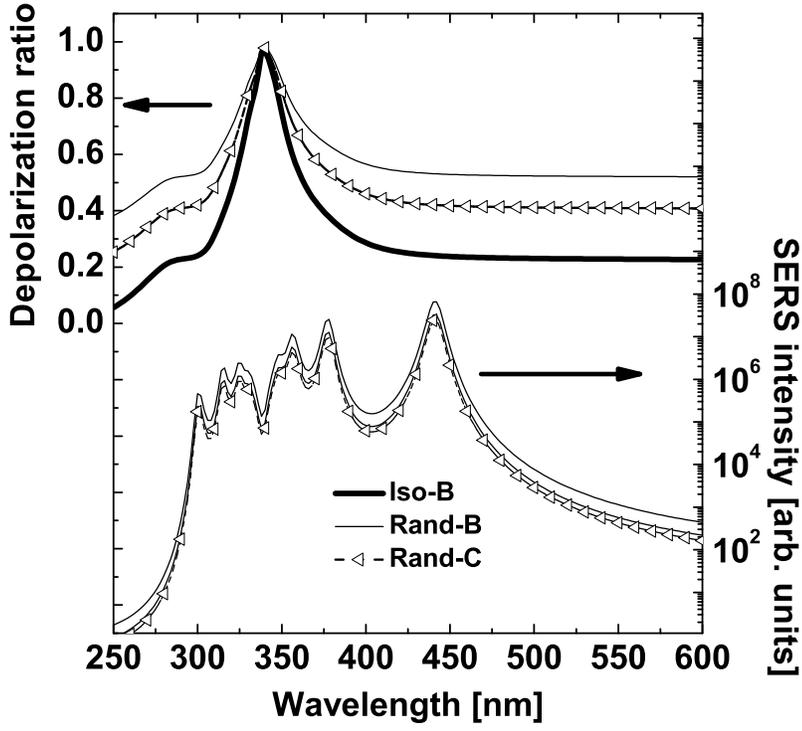}
 }
\caption{Depolarization ratios (top) and SERS intensities (bottom)
for $Iso-B$, $Rand-B$, and $Rand-C$, for a 2D dimer with
separation $d=0.1a$. Note that the depolarization ratio is not
affected by all the structures appearing as resonances in the SERS
enhancement.} \label{fig6}
\end{figure}

\begin{figure}[h]
 \centering{
  \includegraphics[height = 12 cm]{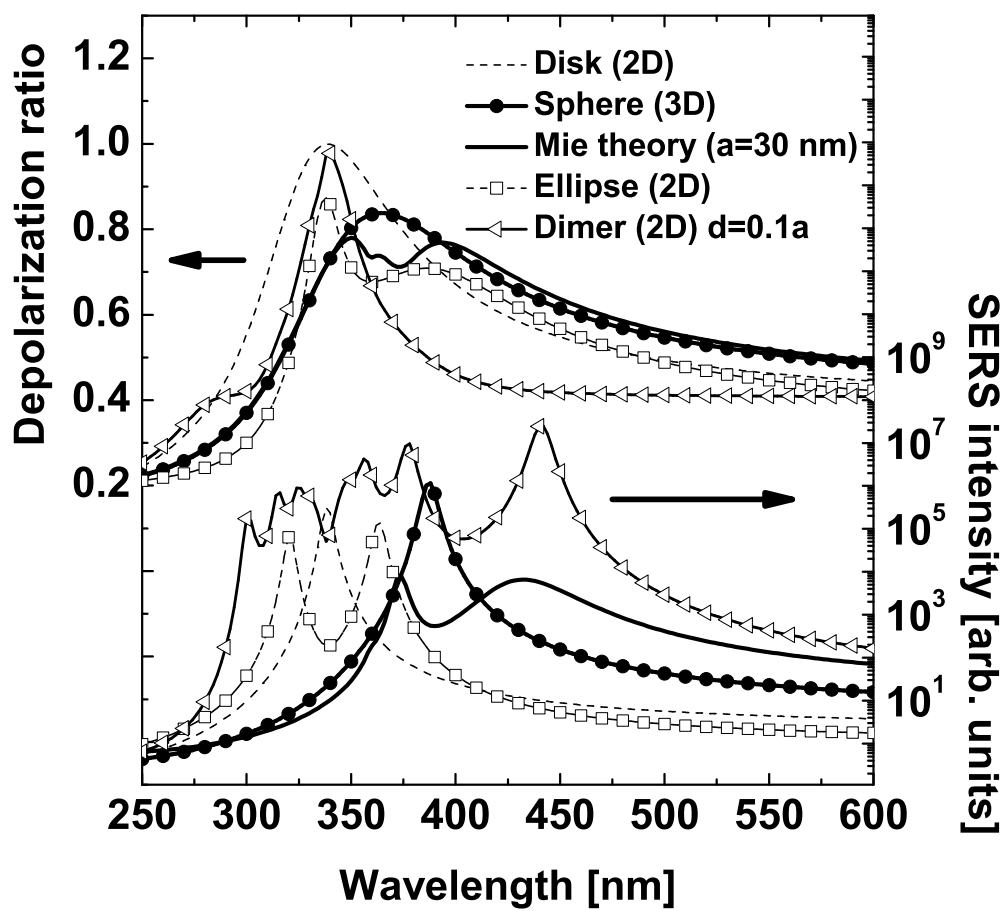}
 }
\caption{Depolarization ratios $\rho$ (top) and SERS intensities
(bottom) for the case $Rand-C$ for different model geometries.
This figure summarizes the various effects on $\rho$ of shape,
dimensionality, retardation, and coupled resonances.} \label{fig7}
\end{figure}

Finally, it is widely believed that the largest SERS enhancements
originate from interactions between two or more particles, leading
to coupled plasmon resonances. In many situations, it is likely
that the SERS signal originates mainly from such interactions. We
would therefore like to enquire whether such coupled resonances
could affect our conclusions on the depolarization ratios. To do
so, we calculated the depolarization ratio in the electrostatics
approximation for two closely spaced 2D disks separated by $d=0.1
a$ (where $a$ is the radius). The results are summarized in Fig.\
\ref{fig6}. The structure of the SERS intensity profile in the
region of the single plasmon resonance (from 300 to 380~nm) is
very complex. The coupled plasmon resonance is substantially
red-shifted to 440~nm in this case (404~nm for a separation of
$d=0.2 a$, not shown). This resonance is also the one exhibiting
the highest integrated SERS intensity, as expected. The complexity
of the intensity profile is due to the averaging of all possible
orientations. In the literature, no orientation averaging is
usually considered, hence the intensity profiles exhibit much less
structure in general. A wide variety of depolarization ratios is
observed depending of the assumption made. Some of the discarded
models in this section (like $Tan-A$, $Tan-B$, and $Tan-C$)
predict values for $\rho$ well above 1, a situation not observed
experimentally. The most surprising result of the inclusion of
coupled plasmons in dimers is that they seem to affect only the
SERS intensity but not so much $\rho$, as can be observed in Fig.
\ref{fig6}. When compared to the results for a single disk (see
Fig.\ \ref{fig7} for example), coupled resonances appear to narrow
the resonance profile of $\rho$ and make it tend faster towards
its limiting long wavelength value. This leads to a small decrease
in $\rho$ for intermediate wavelengths.

Some of the other models not shown in the figure like $Norm-B$ and
$Norm-C$, show a $\rho$ that tends quickly towards its limiting
value of $1/5$, which was also predicted for disks and spheres (2D
and 3D) and ellipses (2D). A value of $\rho=1/5$ should therefore
be observed in most SERS conditions if one of these assumptions
were correct. This is not the case, for experiments with a
particular analyte like BTZ2 (used in the experimental section
later on), but could be correct in other cases. We therefore
conclude that the assumptions $Norm-B$, and $Norm-C$ for the
interaction of the dye with the metal cannot apply for the
specific analyte we shall explore later.

\section{Some experimental evidence}

We have already ruled out a number of possibilities and now
concentrate on the most likely ones in terms of a possible
connection with experimental results, namely: $Iso-B$, $Rand-B$
and $Rand-C$ models. All three cases show a qualitatively similar
behavior, which is summarized in Fig.\ \ref{fig7} for the case
$Rand-C$ for various model geometries. We can draw from the
previous calculations a number of conclusions about the
depolarization ratios in these cases:
\begin{itemize}
\item In a SERS experiment on colloidal solutions with
non-interacting colloids, where single colloids would contribute
to a substantial fraction of the intensity, one should in
principle observe signals from a distribution of spheroidal
colloids, and possibly some elongated ones. One therefore expects
to observe a $\rho$ of a maximum of about 0.7-0.9. However, we saw
in two instances (2D ellipses, and 3D spheres with retardation),
that the maxima of $\rho$ corresponded to dips in the SERS
intensity, and vice-versa. If the excitation wavelength
$\lambda_L$ is within the range of particle resonances, resonant
particles will contribute more to the signal, and a smaller value
for $\rho$ should be observed (we are in the dip associated with a
maximum of SERS intensity). When $\lambda_L$ is increased beyond
most particles resonances, $\rho$ should go through a maximum and
eventually decrease slowly towards a limiting value depending on
the model chosen. However, this limiting value is very difficult
to measure, because the SERS intensity also decreases
substantially.
 \item
In interacting colloid solutions, SERS signals coming from
interacting colloids will overwhelm that coming from single
particles. Since the predicted depolarization ratio of a dimer
vary little with wavelength at the coupled red-shifted resonance
and beyond, one would expect a constant $\rho$ at long
wavelengths. $\rho$ should also increase as $\lambda_L$ is
decreased towards the single (uncoupled) plasmon resonance. Other
scenarios could be considered: for example, because the signal
comes from a collection of various interacting objects, some
interactions may be more dominant at some wavelengths. In a simple
dimer picture, one expects that as the wavelength increases, more
coupled (and therefore more red-shifted) resonances are dominating
the signal. A calculation of 2D dimers with different separations
shows that increased coupling should have a negligible effect on
$\rho$. Another possibility is that the colloids may be able to
merge and form a single elongated particle. Their behavior could
then be more similar to that of an ellipsoidal particle.
\end{itemize}

Far from attempting a full experimental study of the effects
reported here for reasons of space, we give an experimental
indication of some of the important aspects that can be observed
experimentally which do support the idea that depolarization
ratios in SERS are affected by plasmon resonances. Accordingly, we
measured depolarization ratios of SERS active liquids under
various conditions and at different wavelengths.

The experimental details are identical to those reported in Ref.
\cite{Maher}, except for the analyte which in these experiments is
the newly developed benzotriazole dye:
4-(5'-azobenzotriazolyl)naphthalen-1-ylamine (BTZ2)\cite{Smith}.
The samples were prepared by mixing 0.5 mL of silver
colloids\cite{colloids} with 0.5 mL of water or KCl (concentration
10 mM), and adding 40~$\mu$L of 5~$\mu$M BTZ2 to obtain an end
concentration of 0.2~$\mu$M of dye. The samples were then diluted
by a factor of 4 (in water or KCl) to test for a possible effect
of with multiple scattering on the depolarization ratio. No
evidence for this were found when comparing diluted and
non-diluted samples. To calculate depolarization ratios, the
signals were averaged over periods between 5 and 20 minutes, to
avoid problems associated with SERS fluctuations. The ratios is
then estimated for 3 consecutive experiments to check for
reproducibility and estimate errors.

\begin{figure}[h]
 \centering{
  \includegraphics[height = 12 cm]{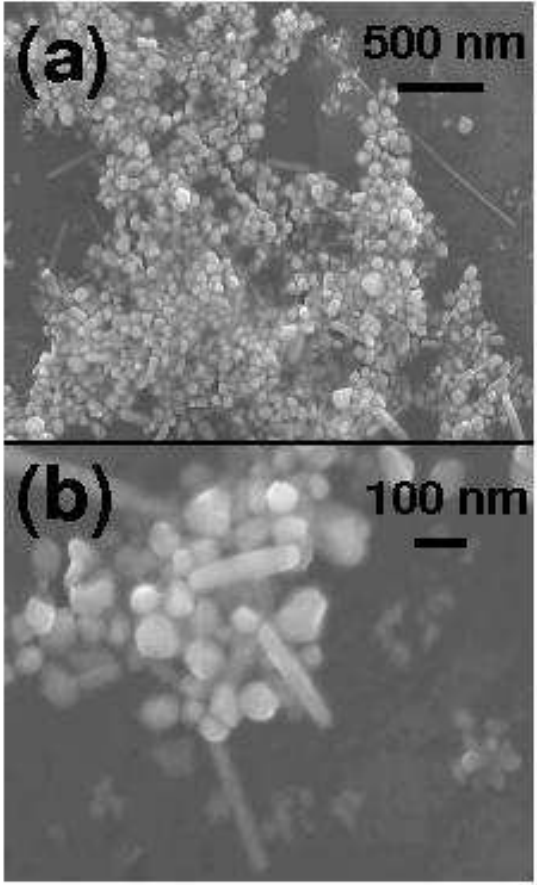}
 }
\caption{Two scanning electron microscope images of Ag-colloids
used in the experiments reported here and prepared by the method
of Ref. \cite{colloids}. The images are at different
magnifications and give an idea of the typical spread in sizes (a)
and the details of individual particles (b).}
 \label{fig8}
\end{figure}

These samples are stable for several weeks at room temperature.
There is substantial experimental evidence of poly-dispersion in
such Ag colloidal solution. Figure \ref{fig8} gives a more
definite idea of the complexity of real colloids, like the ones we
used in the experiments reported below, in terms of size and/or
shape distribution. The colloids have an average radius of
$a=30$~nm and exhibit an absorption spectrum peaking at 454~nm.
Spheres of $a=30$~nm should have a resonance at $430$~nm, and the
observed red-shift in absorption is therefore attributed to
variations in the shape of the particles. Without KCl, there is a
strong coulombic repulsion among particles, and the SERS signal
originates from single particle resonances. The addition of KCl
results in a screening of this repulsive force, making the
occurrence of closely spaced pairs more likely either by
collisions or direct formation of dimers or small clusters. This
is enough to observe a large increase in SERS signal, which shows
that interacting particles dominate the signal. Such samples
remain however stable with no aggregation into large clusters.

\begin{figure}[h]
 \centering{
  \includegraphics[height = 12 cm]{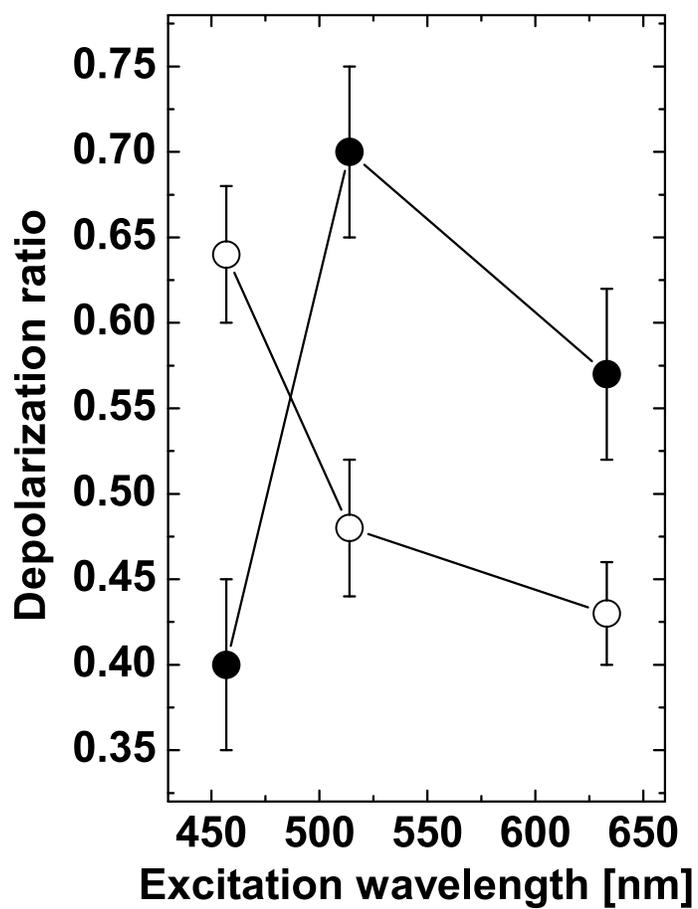}
 }
\caption{Experimental depolarization ratios for 3 different laser
excitations: 457, 514, and 633 nm, for a colloid sample with
(empty circles) and without (solid circles) KCl. The analyte is
the benzotriazole dye BTZ2. These data show a dispersion in laser
wavelength as well as a change depending on the interacting state
of the colloids (KCl concentration).} \label{fig9}
\end{figure}

BTZ2 has several Raman peaks, the most prominent being around
1410~cm$^{-1}$. We measured the depolarization of this peak for a
high dye concentration in water (no colloid) with 514~nm
excitation and obtained $\rho=0.33$, indicating a uniaxial Raman
tensor. The depolarization ratios for this peak measured under
SERS conditions are summarized in Fig. \ref{fig9} as a function of
laser wavelength, for samples without KCl (single particles) and
with KCl (interacting particles). It is interesting to note that
under SERS conditions, the depolarization ratios of all the other
peaks was nearly the same as this one; a situation not happening
in the pure dye and confirming that $\rho$ is mainly determined by
plasmon resonances rather than by the Raman tensor of the peaks.
The background (or continuum, a well-known characteristic of SERS)
also exhibits the same depolarization ratio, indicating its origin
is strongly related to that of the SERS signal. Our experimental
findings are consistent with the previous discussion for this
specific experimental implementation. The models $Rand-C$,
$Rand-B$, or $Iso-B$ show a good qualitative agreement. $Rand-C$
shows the best quantitative agreement, while $Rand-B$ seem to be
predict too high a depolarization ratio at longer wavelength.
$Iso-B$ is not relevant to this specific experiments, since we
know from the depolarization ratio measurement in the pure dye
that the tensor is not isotropic, at least when not adsorbed on
the metal. Note the clear presence in Fig. \ref{fig9} of a
dependence on both the laser excitation and the KCl concentration,
which reveals the role of plasmon resonances brought about by
interactions among colloids. Both effects are expected from the
general considerations presented in this paper. We discuss briefly
the results in the next section.

We believe that further experiments on depolarization ratios could
prove useful to the general understanding of SERS. In particular,
experiments where parallel and perpendicular polarization
intensities can be monitored simultaneously would enable one to
study the fluctuations of $\rho$ with time. This could provide
insight into the origin of the SERS fluctuations, and a test of
their interpretation as coming from single molecules.

\section{Discussion and Conclusions}

The rotationally invariant components of the Raman tensor measured
by depolarization ratios have a limited amount of information
about the symmetry of the vibration. If a new molecule is measured
in liquid, of which no previous information is known, we can infer
the symmetry of the Raman tensor only to a certain extent. In
general a combination of techniques are required. The same
limitations occur for the changes in depolarization ratios by the
presence of plasmon resonances: there is only a limited amount of
information we can infer about the orientation of the molecule on
the surface and it is very likely that we cannot decide between
two competing models in certain cases. Like in the original case
without plasmons, a combination of techniques will be needed for
that task.

Still, what this paper has unequivocally shown is that {\it
plasmon resonances produce "anomalous" depolarization ratios in
standard SERS conditions and that the presence of the plasmons
cannot be neglected or ignored}. The key issue highlighted in this
paper can be summarized with the textbook example of local fields
presented in Fig. \ref{fig2}. SERS is about local fields, and the
polarization of the local field does not follow directly from that
of the incident field. In that respect, we can safely say that
plasmons are the most important factors controlling the
depolarization of the scattering process under SERS conditions and
are responsible for the large number of anomalies seen in
experiments; a large fraction of which have not been explained or
remain unaccounted for. The experimental results of the previous
section do suggest that both the excitation wavelength and the
state of aggregation of the colloids (producing coupled plasmon
resonances) play a role in the depolarization ratios that are
observed experimentally.

In this respect, we also believe that the interpretation of
depolarization ratios put forward in several previous
publications\cite{Brus,Brus2} will probably have to be revised.
Depolarization ratios have been used in the past to infer
something about the orientation of the Raman tensor of a few (or
one) molecules, completely ignoring the fact that the laser and
Stokes field couple to the local field through the local plasmons
and their symmetries. We believe that a re-interpretation of the
data including the role of the plasmons might be necessary in
those cases.

Finally, the limited amount of experimental information we have at
the moment seems to to favor model $C$ with respect to $B$ for the
scattering efficiency. This is a longstanding unresolved issue in
SERS: whether the enhancement operates separately for the laser
and Stokes-field thus producing a factor of
$\propto\left|\vec{E}\right|^4$ in the scattering efficiency, or
whether it is only the enhancement at the laser frequency
$\propto\left|\vec{E}\right|^2$ followed by a re-emission process.
Far from resolving this issue here, we suggest that depolarization
ratios could help to discard possibilities and learn a bit more
about the SERS mechanism at a microscopic level. This is a unique
aspect of SERS provided by the fact that the probe is in direct
contact with the local field.

\section{Acknowledgements}
PGE acknowledges partial support for this work by the Engineering
and Physical Sciences Research Council (EPSRC) of the UK under
grant GR/T06124. Discussions on SERS with M. Cardona
(Max-Planck-Institute, Stuttgart), Lesley Cohen (Imperial
College), and Robert Maher (Imperial College) are gratefully
acknowledged. We are indebted to Richard Tilley and David Flynn
(Victoria University) for help and advice with electron microscopy
and to Mike Dalley for help with the SERS measurements.

\newpage

\end{document}